\documentclass[aps,floatfix,twocolumn,a4paper,showpacs, nofootinbib,superscriptaddress,10pt]{revtex4}
\usepackage{graphicx,float}\usepackage{graphicx,float}
\usepackage[all]{xy}
\usepackage{amsmath,upgreek}
\usepackage{diagbox}
\usepackage{amssymb}
\usepackage{color}
\usepackage{epsfig,bm}		
\usepackage{graphicx,epstopdf}
\usepackage{subfigure}
\usepackage{pdfpages,slashed}
\usepackage[colorlinks,hyperindex]{hyperref}

\definecolor{green1}{RGB}{0,128,0}
\hypersetup{backref=true,pagebackref=true}
\hypersetup{%
  colorlinks = true,
  linkcolor  = blue,
  citecolor = cyan,
}
\usepackage{bookmark,textgreek}
\usepackage{hyperref,color,xcolor}
\hypersetup{hidelinks,hyperindex=true,colorlinks=true,breaklinks=true,urlcolor= blue}
\hypersetup{%
  colorlinks = true,
  linkcolor  = blue
}\usepackage{amssymb}
\newsavebox{\foobox}

\usepackage{graphicx,float,tikz}
\usepackage[all]{xy}
\newcommand\ringring[1]{%
  {% make an Ord atom
   \mathop{\kern0pt #1}\limits^{% set a box over the variable
     \vbox to-1.85ex{
       \kern-2ex % lower the ring accents
       \hbox to 0pt{\hss\normalfont\kern.1em \r{}\kern-.45em \r{}\hss}%
       \vss % fill
     }% end of \vbox
   }% end of the superscript
  }% end of \mathop
}\newcommand\orcidroldao{{\href{https://orcid.org/0000-0003-3978-532X}{\orcidicon}}}
\newcommand{\orcidicon}{%
	\begin{tikzpicture}
	\draw[lime, fill=lime] (0,0)
		circle [radius=0.16]
		node[white] {{\fontfamily{qag}\selectfont \tiny ID}};
	\draw[white, fill=white] (-0.0625,0.095)
		circle [radius=0.007];
	\end{tikzpicture}	\hspace{-2mm}
}
\newcommand{\bpartial}{\mathop{{{\scalebox{0.65}{${{\scalebox{1.3}{$\partial$}}}$}}}\kern -4pt\raisebox{.8pt}{$|$}}}

\newcommand{\bes}{\begin{subequations}}
\newcommand{\ees}{\end{subequations}}
\def\beq{\begin{eqnarray}}
 \newcommand{\clt}{\textcolor{black}}
  
\def\eeq{\end{eqnarray}}
\def\be{\begin{equation}}
\def\ee{\end{equation}}

\begin{document}

\title{Nuclear information entropy, gravitational form factor, and glueballs in AdS/QCD}
\author{G. Karapetyan}
\email{gayane.karapetyan@ufabc.edu.br}
\affiliation{Federal University of ABC, Center of Natural Sciences, Santo Andr\'e, 09580-210, Brazil}
\affiliation{Federal University of ABC, Center of Mathematics, Santo Andr\'e, 09580-210, Brazil}
\affiliation{Perimeter Institute for Theoretical Physics,
Waterloo, Ontario, N2L 2Y5, Canada}

\author{R. da Rocha\orcidroldao\!\!}
\email{roldao.rocha@ufabc.edu.br}
\affiliation{Federal University of ABC, Center of Mathematics, Santo Andr\'e, 09580-210, Brazil}

\begin{abstract}
The nuclear configurational entropy (NCE) is employed to derive two parameters encoding the Pomeron Regge trajectory and the spectrum of closed strings, which are also related to the glueball mass, precisely matching data of the TOTEM collaboration at the LHC. In the Regge regime of holographic QCD, the (Reggeized) glueball propagator and the gravitational form factor of the proton occupy a relevant spot in AdS/QCD. They are used for deriving the cross-sections for the TeV scale proton-proton scattering process, which are the main ingredient for computing the NCE. The nuclear configurational stability is also addressed.
 \end{abstract}
\pacs{89.70.Cf, 11.25.Tq, 12.38.-t}
\maketitle

\section{Introduction}
Measures of information entropy comprise useful techniques to scrutinize any compact physical system that is spatially localized, in particular when extreme conditions set in. They comprise the principal constituent for transmitting data and can be used to quantify information associated with events fitting any probability distribution \cite{Shannon:1948zz,Gleiser:2011di}.
The configurational entropy (CE) represents an important information entropy measure and has been comprehensively utilized in quantum chromodynamics (QCD) to determine the behavior of the particles and resonances at high-energy regimes 
\cite{Bernardini:2016hvx,Bernardini:2018uuy,Ferreira:2019inu,Karapetyan:2018oye}.
To describe properly the theory of strong interactions at high-energy ranges, AdS/QCD was proposed, describing the duality between gravity in an AdS bulk and QCD on the AdS boundary \cite{Brodsky:2014yha,EKSS2005}. In the hard-wall AdS/QCD, glueballs, baryons, and mesons can be described in a codimension-1 AdS bulk with a sharp cut-off along the dimension out of the boundary. On the other hand, the soft-wall AdS/QCD takes into account a dilatonic field inducing a smooth cutoff along the bulk. 
The non-perturbative QCD approach permits studying confinement and mesonic states corresponding to Yang--Mills fields, governed by a gauge flavor theory
\cite{Karch:2006pv,Branz:2010ub,Colangelo:2008us}.
Strongly-coupled QCD theory at low-energy regimes, within AdS/QCD models, imparts an appropriate description of several experimental results.
Recently, a comprehensive investigation of hadron interactions through the AdS/QCD approach and CE techniques has been implemented, matching experimental data and phenomenological aspects as well \cite{daRocha:2021ntm,daRocha:2021imz,Ferreira:2019nkz,Barbosa-Cendejas:2018mng,Braga:2018fyc}.
In some other investigations, AdS/QCD has been employed to scrutinize properties of different light- and heavy-quark mesons as well as tensorial mesonic resonances, baryonic resonances, families of exotic mesons, and glueball fields, under extremal conditions
\cite{Colangelo:2018mrt,Ferreira:2020iry,MarinhoRodrigues:2020yzh,Bernardini:2016qit,Braga:2017fsb,Karapetyan:2018yhm,Karapetyan:2019ran,Ma:2018wtw}.
The outcomes of these developments support and corroborate data at the Particle Data Group (PDG) \cite{pdg} with great accuracy, besides proposing the next generation of several families of mesonic and baryonic resonances to be still probed by running experiments.

Interactions among quarks and gluons, in the high-energy limit, are perturbative in their essence. However, the low-energy regime comprehends a non-perturbative description, that is usually brought into play to study important questions in QCD. Amidst relevant themes, color confinement, formation processes of hadrons out of quarks and gluons -- including the emergence of the quark-gluon plasma and the color string decay into hadrons -- and phase transitions as well play a prominent role in QCD \cite{FolcoCapossoli:2016uns}. An important phase is the color glass condensate formed by saturated gluon matter, with universal features of hadrons studied by the nuclear CE (NCE) \cite{Karapetyan:2021vyh,Karapetyan:2021ufz,Karapetyan:2020yhs}. Usual techniques to investigate non-perturbative QCD involves lattice discretization methods. One can also go into hadronic resonances, encompassing gluonic bound states forming glueball fields, which partakes in interactions among hadrons \cite{Gutsche:2016wix}. 
With appropriate actions ruling the dynamics of massive gluons, glueballs mass spectra can be obtained, matching QCD (lattice) phenomenology, including the soft pomeron in the context of current HERA data \cite{Sergeenko:2011kf}.

The comportment of hadrons at high-energy regimes, as well as their mutual interactions, remains one of the most coveted phenomena to investigate in QCD.
The perturbative QCD approach does not allow the appraisal of the main characteristics of the proton-proton ($pp$) and proton-antiproton ($p \bar p$) reactions, as the energy of scattering processes is exceedingly high.
Taking the classical Regge regime \cite{Xie:2019soz}, inelastic hadron-hadron scattering cross-sections can be derived through Mandelstam variables $s$ and $t$, where $|t|\ll s$ and $|t|$ is assumed to be upper bounded by the QCD characteristic scale of confinement.
The cross section amplitude is calculated as $\mathcal M (s,t) \sim s^{\upalpha(t)}$ where $\upalpha(t) = \upalpha(0) + \upalpha^{\prime}t$ is the Regge trajectory 
with constants $\upalpha(0)$ and $\upalpha^{\prime}$. For mesons, experimental data yield $
J= \upalpha(0) + \upalpha^{\prime} m^2$,
for $J$ denoting the meson spin and $m$ the mass of the meson. It underlies the interpretation of hadron scattering through an infinite channel of exchanging mesons.
To unfold and get across current data describing cross-sections, elastic scattering, and diffraction dissociation measurements at the LHC (TOTEM \cite{TOTEM:2017asr,TOTEM:2012oyl}), a complete model is required.
 There is one common point at $\upalpha(0) \lesssim 0.6$, for which mesonic trajectories intercept. This fact can be better described by a theory of the Pomeron.
Within this setup, the leading Pomeron trajectory cuts off at the value of
$\upalpha(0)$ that is just slightly higher than unit, and is called the soft Pomeron intercept, which it is impossible to deduce from pure QCD.
In AdS/QCD, the Pomeron can be successfully addressed \cite{FolcoCapossoli:2019imm,Rodrigues:2016cdb,FolcoCapossoli:2015jnm,Boschi-Filho:2005xct,Ballon-Bayona:2017vlm}. AdS/QCD successfully characterizes fundamental features of elastic $pp$ and $p \bar p$ scattering, in particular in the Regge regime.
The Pomeron exchange regulates the $pp$ scattering through the Reggeized glueball propagator and the associated proton (gravitational) form factor.
Besides, in the context of string theory, glueballs can be described by closed strings \cite{Hu:2017iix}. The coupling between the proton and the Pomeron was applied to derive cross sections for $pp$ scattering. To implement it,
the gravitational form factor can be derived from the AdS/QCD model 
\cite{Xie:2019soz,Domokos:2009hm}.

The energy density in the standard CE encrypts the main information about the physical system. 
Instead, when the main spatially-localized, Lebesgue-integrable function is the cross-section, the NCE is, therefore, more appropriate for studying QCD. 
The main aim of this work is to use the NCE to obtain two parameters that regulate the Pomeron Regge trajectory and the spectrum of closed strings, which are also related to the glueball mass, showing that they match TOTEM data with great accuracy. In Section \ref{sec1} the proton-Pomeron coupling is represented for $pp$ scattering in the Regge regime, and the Pomeron exchange is derived through the amplitude of the $2^{{\scalebox{0.55}{++}}}$ glueball exchange. Using the proton gravitational form factor in AdS/QCD, in Section \ref{3} the main results for the NCE computed upon the $pp$ differential cross-sections are derived, discussed, and analyzed.
Section \ref{con} comprises further discussions and final comments.

\section{Holographic \emph{pp} scattering and proton gravitational form factor in AdS/QCD}
\label{sec1}

The process of elastic $ pp$ scattering can be well represented within the Pomeron exchange model, for the scattering amplitude being deduced from the Pomeron propagator and the gravitational form factor of the incident proton.
It is worth stressing some important aspects of the model. The proton and the Pomeron are correlated due to the
lowest state on the leading Pomeron trajectory regarding the $2^{{\scalebox{0.55}{++}}}$ glueball field. Also, the glueball propagator comprises a quantum superposition of gluonic states, however just counting terms whose amplitude are leading logarithmic, to embrace the entire set of field states constituting the Pomeron trajectory.
The Pomeron can be portrayed by taking into account the $2^{{\scalebox{0.55}{++}}}$ glueball exchange amplitude.
The QCD stress-energy-momentum tensor, ${{\scalebox{0.95}{$\textsc{T}$}}}_{\mu \nu}$, is used along with the
$2^{{\scalebox{0.55}{++}}}$ glueball field, here represented by a symmetric traceless tensor $h_{\mu\nu}$ to constitute the action 
\cite{Xie:2019soz}
\begin{equation}
\label{e1}
S = \uplambda \int h^{\mu \nu}{{\scalebox{0.95}{$\textsc{T}$}}}_{\mu \nu}\,d^4 x,
\end{equation}
where $\uplambda$ quantifies the coupling between the tensors $h_{\mu\nu}$ and the QCD stress-energy-momentum tensor ${{\scalebox{0.95}{$\textsc{T}$}}}_{\mu \nu}$. It has been obtained experimentally using TOTEM data, as $\uplambda = 9.95\pm0.12$ GeV${}^{-1}$.
The proton and the glueball are here encoded in AdS/QCD by the
fact that dual gravity in the AdS bulk is implemented by the graviton, which transforms as the $2^{{\scalebox{0.55}{++}}}$ glueball. %To encompass hadronic fields, one usually takes a Yang--Mills gauge field that is dual to the axial-vector current density operator that originates pions in holographic QCD.

\clt{Denoting by $s$ the helicity}, the proton-glueball-proton vertex is represented by 
\cite{Domokos:2009hm}
\begin{widetext}
\begin{equation} \label{e0}
\begin{split}
\langle p', s'|{{\scalebox{0.95}{$\textsc{T}$}}}_{\mu \nu} (0) | p, s \rangle = %
\bar u(p', s')\left[\frac{A (t)}{2} \upgamma_{(\mu} P_{\nu)}
\!+\! iB(t) \frac{P_{(\mu} \upsigma_{\nu) \rho} k^\rho}{4m_p} \!+\! C(t) \frac{(k_\mu k_\nu - \upeta_{\mu \nu } k^2)}{m_p}\right] u(p, s), %
\end{split}
\end{equation}
\end{widetext}
where, the Dirac spinor $u(p,s)$ is governed by the Dirac equation $(\clt{\gamma^\mu p_\mu}-m\mathbb{I})u(p,s)=0$, with spin sum \beq
\clt{\sum_s {u}(p,s)\bar{u}(p,s)={\gamma^\mu p_\mu+m\mathbb{I}}},
%Eq. 3.112, A Modern Introduction to Quantum Field Theory
%Michele Maggiore, 2005
\eeq
and
\beq
k_\mu = (p' - p)_\mu,\qquad t = k^2,\qquad P_\mu = \frac12(p + p')_\mu,\eeq for $m_p^2=p^2=p^{\prime2}$ denoting the square of 
the proton mass, whereas $A(t)$, $B(t)$, and $C(t)$ encode the mechanical form
factors of the proton interaction.
These parameters contribute in different ways to the value of scattering amplitude at $ |t|\ll s$, whereas
$C(t)$ is usually suppressed by a $|t|/s$ factor, whereas $B(t)/A(t)\ll 1$. The term $A(t)$ is known as the gravitational form factor and two main limits are regarded
\beq
\lim_{t\to0}A(t)=1,\qquad \lim_{t\to0}B(t)=0.\eeq
 The scattering processes $ pp$ and
$ p \bar p$ will be considered, in which $2^{{\scalebox{0.55}{++}}}$ glueball exchanges are realized in the glueball-proton-proton system. One denotes by $\{p_1, p_2\}$ the ingoing momenta and $\{p_3, p_4\}$ the outgoing momenta, being $k = p_2 - p_4 = p_1 - p_3$ the glueball momentum.
The $t$-channel is effectively the only channel in the Regge regime, wherein the $2^{{\scalebox{0.55}{++}}}$ glueball propagator reads \cite{Xie:2019soz}
\begin{equation}
\label{e2}
\frac{d_{\upalpha\upbeta\upgamma\updelta}(k)} %
{k^2 - m_{{\scalebox{0.65}{\textsc{g}}}}^2},
\end{equation}
where $m_{{\scalebox{0.65}{\textsc{g}}}}$ denotes the glueball mass and the $\upalpha$, $\upbeta$ are
 Lorentz indexes regarding incoming particles, whereas $\upgamma$ and $\updelta$ denote outgoing ones, and
 {widetext}
\beq
d_{\upalpha \upbeta \upgamma \updelta} &=&
\frac{1}{2}\upeta_{\upalpha (\upgamma} \upeta_{\upbeta \updelta)}
- \frac{1}{2m_{{\scalebox{0.65}{\textsc{g}}}}^2}\left(k_{\upalpha} k_{(\updelta|} \upeta_{\upbeta |\upgamma)}
+ k_{\upbeta} k_{(\updelta|} \upeta_{\upalpha |\upgamma)}\right)\nonumber\\&&+ \frac{24 k^4}{m_{{\scalebox{0.65}{\textsc{g}}}}^4}-
\frac{8k^2}{m_{{\scalebox{0.65}{\textsc{g}}}}^2}
- \frac14 \upeta_{\upalpha \upbeta} \upeta_{\upgamma \updelta}+ \frac{2k_{\upalpha}k_{\upbeta}k_{\upgamma}k_{\updelta}}{3 m_{{\scalebox{0.65}{\textsc{g}}}}^4} \nonumber\\ %
&&+ \frac{3 m_{{\scalebox{0.65}{\textsc{g}}}}^2-k^2}{6 m_{{\scalebox{0.65}{\textsc{g}}}}^4}(k_{\upalpha}k_{\upbeta}\upeta_{\upgamma \updelta}
+ k_{\upgamma} k_{\updelta} \upeta_{\upalpha \upbeta}) .
\eeq
When one takes into account the form factors in Eq. (\ref{e0}) and the propagator (\ref{e2}), the scattering amplitude reads
\begin{eqnarray}\label{e3}
{{\scalebox{0.95}{$\mathcal{M}$}}}_{{\scalebox{0.65}{\textsc{g}}}} & =& \frac{\uplambda^2 A^2 (t)}{8 (t - m_{{\scalebox{0.65}{\textsc{g}}}}^2)} \left[ 2 s (\bar{u}_2 \upgamma_{\upalpha} u_4)(\bar{u}_1 \upgamma^{\upalpha} u_3)\right.\nonumber\\ &&\left.+ 4 p_2^{\upalpha} (\bar{u}_1 \upgamma_{\upalpha} u_3)p_1^{\upbeta} (\bar{u}_2 \upgamma_{\upbeta} u_4) \right].
\end{eqnarray}
Hence the cross section for the exchange of the $2^{{\scalebox{0.55}{++}}}$ glueball field can be expressed as
\begin{equation}\label{diff}
\upsigma_{\scalebox{0.55}{$\textsc{Diff}$}} = \frac{1}{16 \pi s^2} | {{\scalebox{0.95}{$\mathcal{M}$}}}_{{\scalebox{0.65}{\textsc{g}}}} (s, t)|^2
= \frac{\uplambda^4 s^2 A^4 (t)}{16 \pi (t - m_{{\scalebox{0.65}{\textsc{g}}}}^2)^2},
\end{equation}
taking the spin sum in (\ref{e3}). The term $s^2$ in the numerator of Eq. (\ref{diff}) reflects the expected $s^J$ dependence arising from the $2^{{\scalebox{0.55}{++}}}$ glueball field exchange.

Now one can additionally consider string holographic dual models to QCD, wherein mesons [glueballs] are dual states to open [closed] strings. 
Let us remember that the Mandelstam variables are given by 
\begin{subequations}
\beq
s&=&(p_{1}+p_{2})^{2}=(p_{3}+p_{4})^{2},\label{s1}\\
 t&=&(p_{1}-p_{3})^{2}=(p_{4}-p_{2})^{2},\label{s2}\\
 u&=&(p_{1}-p_{4})^{2}=(p_{3}-p_{2})^{2}. \label{s3}
\eeq
\end{subequations}
In the scenario of closed strings, the Virasoro--Shapiro scattering amplitude is given by \cite{Domokos:2009hm}
\begin{widetext}
\begin{equation}
\begin{split}
\mathcal M_{{\scalebox{0.65}{$c$}}}(p_1,p_2,p_3,p_4) & = %
\frac{K_{{\scalebox{0.65}{$c$}}} (p_1, p_2, p_3, p_4)\Upgamma[-a_{{\scalebox{0.65}{$c$}}} (t)] \Upgamma[-a_{{\scalebox{0.65}{$c$}}} (u)] \Upgamma[-a_{{\scalebox{0.65}{$c$}}} (s)]}{\Upgamma[-a_{{\scalebox{0.65}{$c$}}} (t) - a_{{\scalebox{0.65}{$c$}}}(s)]\Upgamma[-a_{{\scalebox{0.65}{$c$}}} (t) - a_{{\scalebox{0.65}{$c$}}} (u)] \Upgamma[-a_{{\scalebox{0.65}{$c$}}} (u) - a_{{\scalebox{0.65}{$c$}}} (s)]},
\end{split}
\end{equation}
\end{widetext}
where $K_{{\scalebox{0.65}{$c$}}}$ stands for a kinematic pre-factor that enhances the cross-section, 
and \beq\label{pom1}
a_{{\scalebox{0.65}{$c$}}}(x)=a_{{\scalebox{0.65}{$c$}}}(0) + a_{{\scalebox{0.65}{$c$}}}'x\eeq is derived when the closed strings spectrum is taken into account. In fact, the quantum angular momentum of particle states that undergo exchanging processes nearby any $t$-channel pole reads $J = \upalpha(t)$. Therefore for the open string setup, the Pomeron trajectory parameters $a_{{\scalebox{0.65}{$c$}}} (0)$ and $a^{\prime}_{{\scalebox{0.65}{$c$}}}$ can be written as \cite{Xie:2019soz}
\begin{equation}
a_{{\scalebox{0.65}{$c$}}} (0) = \frac12\upalpha_{{\scalebox{0.65}{$c$}}} (0)-1, \qquad a_{{\scalebox{0.65}{$c$}}}' = \frac12\upalpha_{{\scalebox{0.65}{$c$}}}'.\label{pom}
\end{equation}
It is worth mentioning that $\frac{a_{{\scalebox{0.65}{$c$}}}(0)}{a_{{\scalebox{0.65}{$c$}}}^\prime}=-\sqrt{m_{{\scalebox{0.65}{\textsc{g}}}}}$.

In the Regge regime, the $u$-dependence (\ref{s3}) can be always encoded in the other Mandelstam variables $t$ and $s$ in (\ref{s1}, \ref{s2}), in such a way that 
\begin{equation} \label{e4}
\mathcal M_{{\scalebox{0.65}{$c$}}}^{{\scalebox{0.65}{\textsc{Regge}}}} = e^{- i \pi a_{{\scalebox{0.65}{$c$}}} (t)} (a'_{{\scalebox{0.65}{$c$}}} s)^{2 a_{{\scalebox{0.65}{$c$}}} (t)} \frac{\Upgamma[-a_{{\scalebox{0.65}{$c$}}} (t)]}{\Upgamma[a_{{\scalebox{0.65}{$c$}}} (t) - \upchi]} K_{{\scalebox{0.65}{$c$}}} (p_1, p_2, p_3, p_4),
\end{equation}
where for the simplest case of $2 \to 2$ scattering of particles with equal mass $m$,
\beq
\upchi = a_{{\scalebox{0.65}{$c$}}} (s) + a_{{\scalebox{0.65}{$c$}}} (t) + a_{{\scalebox{0.65}{$c$}}} (u) = 4 a'_{{\scalebox{0.65}{$c$}}} m^2 + 3 a_{{\scalebox{0.65}{$c$}}} (0),\eeq where $m$ is the mass of the interacted particles.
From Eq. \eqref{e4} one can read off the poles at $a_{{\scalebox{0.65}{$c$}}} (t) = n$ for $n\in\mathbb{N}$, having residue proportional to $\sim s^{2 n}$ that obeys the $t$-channel exchange associated with a Regge trajectory with even angular momentum.
For the lowest state, the exchange amplitude reads \cite{Xie:2019soz,Domokos:2009hm}
\begin{equation}\label{e5}
{{\scalebox{0.95}{$\mathcal{M}$}}} \approxeq \frac{- s^{2} \mathsf{f}(\upepsilon_i)} %
{a_{{\scalebox{0.65}{$c$}}}' \Upgamma[-\upchi] (t - m_{{\scalebox{0.65}{\textsc{g}}}}^2)},
\end{equation}
with polarization functions of the scattered particles $h(\upepsilon_i)$ related to the kinematic pre-factor by
\beq
K_{{\scalebox{0.65}{$c$}}}(p_1, p_2, p_3, p_4) \sim \mathsf{f}(\upepsilon_i) s^k
\eeq
The functions $\mathsf{f}(\upepsilon_i)$ cancel out at the end of the calculation.
The expression of the amplitude in the Regge limit can be written as 
\cite{Xie:2019soz,Domokos:2009hm}
\begin{equation}\label{e6}
\!\!\!\!\!{{\scalebox{0.95}{$\mathcal{M}$}}} (s, t)\sim\frac{ \Upgamma \left[1 \!-\! %
\frac{\upalpha_{{\scalebox{0.65}{$c$}}} (t)}{2}\right]e^{-\frac{ i \pi \upalpha_{{\scalebox{0.45}{$c$}}} (t)}{2}}}{\Upgamma \left[\frac{\upalpha_{{\scalebox{0.65}{$c$}}} (t)}{2} \!-\!1 \!-\! \upchi\right]} \left(\frac{\upalpha'_{{\scalebox{0.65}{$c$}}} s}{2} \right)^{\upalpha_{{\scalebox{0.65}{$c$}}} (t) \!-\! 2} \!\!\!\!\!s^{2}f (\upepsilon_i).
\end{equation}
One can hence correlate the amplitude \eqref{e6} to the one in \eqref{e5}
when substituting the glueball exchange factor
by a (Reggeized) Pomeron propagator,
\beq
\frac{1}{t \!-\! m_{{\scalebox{0.65}{\textsc{g}}}}^2}\mapsto -\frac{\upalpha_{{\scalebox{0.65}{$c$}}}^\prime}{2}e^{-\frac{ i \pi \upalpha_{{\scalebox{0.45}{$c$}}} (t)}{2}}
\frac{\Upgamma[-\upchi]\Upgamma\left[1 \!-\!
\frac{\upalpha_{{\scalebox{0.65}{$c$}}} (t)}{2}\right]}{\Upgamma\left[\frac{\upalpha_{{\scalebox{0.65}{$c$}}} (t)}{2} \!-\!1 \!-\! \upchi \right]} \!\!\left(\frac{\upalpha'_{{\scalebox{0.65}{$c$}}} s}{2} \right)^{\!\!\upalpha_{{\scalebox{0.65}{$c$}}} (t) \!-\! 2}
\eeq
yielding to an invariant amplitude
\begin{equation}
\begin{split}
\!\!\!{{\scalebox{0.95}{$\mathcal{M}$}}} & \!=\! s \uplambda^2 A^2 (t) e^{-\frac{ i \pi \upalpha_{{\scalebox{0.45}{$c$}}} (t)}{2}}
\frac{\Upgamma[-\upchi]\Upgamma\left[1 \!-\!
\frac{\upalpha_{{\scalebox{0.65}{$c$}}} (t)}{2}\right]}{\Upgamma\left[\frac{\upalpha_{{\scalebox{0.65}{$c$}}} (t)}{2} \!-\!1 \!-\! \upchi \right]} \!\left(\frac{\upalpha'_{{\scalebox{0.65}{$c$}}} s}{2} \right)^{\!\!\upalpha_{{\scalebox{0.65}{$c$}}} (t) \!-\! 1}.
\end{split}\label{sss}
\end{equation}
Therefore the differential cross section \eqref{diff} for the exchange of the $2^{{\scalebox{0.55}{++}}}$ glueball field reads
\beq\label{cs}
\upsigma_{\scalebox{0.55}{$\textsc{Diff}$}}=\frac{\uplambda^4 A^4 (t)\Upgamma^2[-\upchi]\Upgamma^2\left[1 -
\frac{\upalpha_{{\scalebox{0.65}{$c$}}} (t)}{2}\right]}{16 \pi \Upgamma^2\left[\frac{\upalpha_{{\scalebox{0.65}{$c$}}} (t)}{2} -1 - \upchi \right]}\left(\frac{\upalpha'_{{\scalebox{0.65}{$c$}}} s}{2} \right)^{2\upalpha_{{\scalebox{0.65}{$c$}}} (t) - 2}\eeq
containing the Pomeron contribution to the $pp$ scattering process.

Therefore it is peremptory calculating the gravitational form factor $A(t)$ in Eq. (\ref{sss}) to compute the NCE of the system. It can be derived in the AdS/QCD framework, when a massive fermionic field in the AdS bulk space is studied, governed by the action 
\cite{Xie:2019soz}
\begin{widetext}
\begin{equation}
\begin{split}
S_\Uppsi& = %
\int \sqrt{g} e^{- \Upphi (z)} \bigg( \frac{i}{2} \bar{\Uppsi} e^P_B \upgamma^B D_P \Uppsi - \frac{i}{2} (D_P \Uppsi)^\dagger \upgamma^0 e^P_B \upgamma^B \Uppsi - (M + \Upphi (z)) \bar{\Uppsi}\Uppsi \bigg) d^5 x,
\end{split}
\end{equation}
\end{widetext}
with the tetrad involving only the $z$ bulk coordinate\footnote{The extra-dimensional coordinate $z$ is an scale energy of QCD.} as $e^P_B = z \updelta^P_B$, and the covariant derivative reads $D_P = {{\scalebox{0.65}{${{\scalebox{1.3}{$\partial$}}}$}}}_P + \frac{1}{8} \upomega_{PAB}[\upgamma^A, \upgamma^B] - i V_P$, where $V_P$ denotes the components of a gauge field and the gamma matrices satisfy the Clifford relation $\{ \upgamma^A,\upgamma^B \} = 2 \upeta^{AB}$. The spin-connection is given by
\beq
\upomega_{\mu\nu z}=\frac{1}{z}\upeta_{\mu\nu},
\eeq
with all other components equal to zero, and $\upgamma^B=(\upgamma^\mu, -i\upgamma^5)$.
The quadratic dilatonic field $\Upphi (z) = (\upkappa z)^2$ is included into the mass term in the action.
The equations of motion for the Dirac fermion read
\begin{equation}
\biggl[ i e^P_B \upgamma^B D_P - \frac{i}{2} ({{\scalebox{0.65}{${{\scalebox{1.3}{$\partial$}}}$}}}_P \Upphi) e^P_B \upgamma^B %
 - (M + \Upphi (z))\biggr] \Uppsi = 0.
\end{equation}
One denotes hereon by \beq
\Uppsi_{{\scalebox{0.6}{\textsc{R, L}}}} = \frac12(\mathbb{I} \pm \upgamma^5) \Uppsi
\eeq
the right- and left-handed chiral projector spinor field components.
In momentum space the Dirac fermion reads
\begin{equation}
\Uppsi_{{\scalebox{0.6}{\textsc{R, L}}}} (p, z) = f_{{{\scalebox{0.6}{\textsc{R, L}}}}}(p, z)\,z^\Updelta\,\mathring\Uppsi(p)_{{\scalebox{0.6}{\textsc{R, L}}}},
\end{equation}
where the $f_{{\scalebox{0.6}{\textsc{R, L}}}}(p, z)$ are bulk-boundary 2-point correlators and the QCD boundary fields are denoted by $\mathring\Uppsi_{{\scalebox{0.6}{\textsc{R, L}}}}$. The left-handed component $\mathring\Uppsi_{{\scalebox{0.6}{\textsc{L}}}} (p)$ sources the baryon bulk operator $\mathcal{O}_{{\scalebox{0.6}{\textsc{R}}}}$, whereas the conformal dimension $\Updelta$ is governed by $f_{{\scalebox{0.6}{\textsc{L}}}} (p, \varepsilon ) = 1$ at the ultraviolet (UV) boundary $\upepsilon$ 
\cite{Xie:2019soz}.

The link between the left- and the right-handed components is paved by the Dirac equation in the boundary,
\beq
\slashed{p}\,\mathring\Uppsi_{{\scalebox{0.6}{\textsc{R}}}} (p) = p\mathring\Uppsi_{{\scalebox{0.6}{\textsc{L}}}} (p).\eeq
Hence, the Dirac equation ruling the fermionic field, for the fermion mass $M = \frac{3}{2}$ can be split into a system of two coupled equations
\begin{eqnarray}
\left[{{\scalebox{0.65}{${{\scalebox{1.3}{$\partial$}}}$}}}_z \!-\! \frac{1}{z}(2 \!-\! \Updelta \!+\!M\!+\! 2 \Upphi)\right] f_{{\scalebox{0.6}{\textsc{R}}}}(p, z) &\!=\!&-pf_{{\scalebox{0.6}{\textsc{L}}}}(p, z), \\ %
\left[{{\scalebox{0.65}{${{\scalebox{1.3}{$\partial$}}}$}}}_z \!-\! \frac{1}{z}(2 \!-\! \Updelta \!-\!M) \right] f_{{\scalebox{0.6}{\textsc{L}}}} (p, z)&=&pf_{{\scalebox{0.6}{\textsc{R}}}}(p, z).
\end{eqnarray}
The kinetic part of the vector field can be written as:
\begin{equation}
S_{{\scalebox{0.6}{\textsc{vector}}}} = \int e^{-\Upphi} \sqrt{g} %
{\rm Tr} \left( - \frac{F_{M N}F^{M N}}{2 g^2_5} \right)d^5 x,
\end{equation}
in which $F_{M N} = {{\scalebox{0.65}{${{\scalebox{1.3}{$\partial$}}}$}}}_{[M} V_{N]}$ and $g_5^2={4\pi^2\,N_f}/{N_c}$ regards the color number-to-flavor number ratio.
When the bulk-boundary 2-point correlator satisfies $V (p, \varepsilon) = 1$, the boundary components the vector field read
\beq
V_\mu (p, z) = V (p, z) V^0_\mu (p),
\eeq where $V^0_\mu (p)$ is associated to the boundary current density operator, $J^V_\mu$.
Thereby, using the gauge $V_z = 0$, the equation of motion has the form 
\begin{equation} 
\left[{{\scalebox{0.65}{${{\scalebox{1.3}{$\partial$}}}$}}}_z^2-\left(z\Upphi'(z)+1\right){{\scalebox{0.65}{${{\scalebox{1.3}{$\partial$}}}$}}}_z+zp^2\right] %
V (p, z) = 0.
\end{equation}
This solution has eigenvalues that give rise to a Kaluza--Klein mass spectrum $p^2 = M^2_n = 4 \upkappa^2 (n + 1)$ for the vector field, for $n\in\mathbb{N}$, in the soft-wall AdS/QCD.

The stress-energy-momentum tensor determines, for fermions, three form factors of Eq. \eqref{e0}, which can be derived by the 3-point function
\begin{equation}
 \label{e9}
\left\langle\, 0\, \big|\mathcal{O}^i_{{\scalebox{0.6}{\textsc{R}}}} (x_1) %
{{\scalebox{0.95}{$\textsc{T}$}}}_{\mu\nu}(x_2) \bar{ \mathcal{O}}^j_{{\scalebox{0.6}{\textsc{R}}}} (x_3) \big|\, 0\, \right\rangle,
\end{equation}
with AdS bulk metric in Poincar\'e coordinates
\begin{equation}
ds^2 = g_{AB} dx^A dx^B = %
\frac{1}{z^2} \left(\upeta_{\mu \nu} dx^\mu dx^\nu - dz^2 \right).
\end{equation}
One can emulate the stress-energy-momentum tensor operator by perturbing the metric,
\beq\label{e20}
\upeta_{\mu \nu} \mapsto \upeta_{\mu \nu} + h_{\mu \nu}.
\eeq
Therefore the action in the second-order approximation for the perturbation reads
\begin{equation}
S_{{\scalebox{0.65}{\textsc{gravity}}}} = %
- \int \frac{e^{- (\upkappa z)^2}}{4 z^3} %
\left[ ({{\scalebox{0.65}{${{\scalebox{1.3}{$\partial$}}}$}}}_zh_{\mu \nu})({{\scalebox{0.65}{${{\scalebox{1.3}{$\partial$}}}$}}}_z{h^{\mu \nu}}) %
+ h^{\mu \nu} \Box h_{\mu \nu} \right]\,d^5 x,
\label{e10}
\end{equation}
in the transverse and traceless gauge. 
Fluctuations of the AdS bulk background metric, coupling to the stress-energy-tensor for the matter fields, mimics Eq. (\ref{e1}),
\begin{equation}
\label{e11}
S =\frac12\int h^{AB}{{\scalebox{0.95}{$\textsc{T}_{AB}^{{\scalebox{0.65}{\textsc{matter}}}}$}}}\,d^5 x,
\end{equation} where ${{\scalebox{0.95}{$\textsc{T}_{AB}^{{\scalebox{0.65}{\textsc{matter}}}}$}}}$ encodes matter fields representing a soliton solution describing the proton \cite{Xie:2019soz}. The action (\ref{e10}) yields the Einstein's field equation,
\begin{equation}
\left[ {{\scalebox{0.65}{${{\scalebox{1.3}{$\partial$}}}$}}}_z \left(\frac{e^{- (\upkappa z)^2}}{z^3} {{\scalebox{0.65}{${{\scalebox{1.3}{$\partial$}}}$}}}_z \right) %
+ \frac{e^{- (\upkappa z)^2}}{z^3} p^2 \right] h(p, z) = 0,
\label{e12}
\end{equation}
whose solution employs the Kummer's confluent hypergeometric function,
\begin{equation}
\begin{split}
h (p, z)\Big\vert_{p^2= - Q^2} &= \Upgamma\left(\frac{Q^2}{4 \upkappa^2} + 2\right) U \left(\frac{Q^2}{4 \upkappa^2}, - 1; (\upkappa z)^2 \right)\end{split}
\label{eq:b-to-b_propagator}
\end{equation}
The quadratic dilaton parameter $\upkappa = 350$ MeV depends on the proton mass and the $\rho$ meson mass. Still,
to obtain the gravitational form factor $A(t)$ from 
the 3-point function \eqref{e9}, one can superpose the solutions for the Dirac fermionic field and for the metric perturbation, as
\begin{equation}\label{pgf}
A (Q) =
\int
\left(\Uppsi_{{\scalebox{0.6}{\textsc{L}}}}^2 (z) + \Uppsi_{{\scalebox{0.6}{\textsc{R}}}}^2 (z) \right) \frac{e^{- (\upkappa z)^2}}{2 z^{2 M}} h (p, z)\Big\vert_{p^2= - Q^2} \,dz.
\end{equation}
This expression can be forthwith replaced in the differential cross-section (\ref{cs}), to compute the NCE in the next section.

\section{The nuclear configurational entropy for the glueball exchange with Pomeron contribution to \emph{pp} scattering}
\label{3}
In the NCE setup, it is possible to realize the splitting of the differential cross-section for proton-proton scattering into weighted wave mode components.
Hence, as the differential cross-section in \eqref{cs} is written already in the momentum space, regarding the Mandelstam variables, the nuclear modal fraction can be immediately read off the differential cross-section, encoding the energy-weighted correlation. In particular, as expressed in Eq. (\ref{diff}), the differential cross-section
is a spatially-localized function designated by the squared amplitude. 
The NCE approach is then used to calculate and analyze the parameters
in Eq. (\ref{pom}), related to the Pomeron Regge trajectory and the spectrum of closed strings. The nuclear modal fraction can be obtained as
\begin{equation}\label{modall}
n_{\upsigma(s,t)}=\frac{\left\vert \upsigma_{\scalebox{0.55}{$\textsc{Diff}$}}(s,t) \right\vert^2}{\displaystyle\iint_{\mathbb{R}^2}\left\vert \upsigma_{\scalebox{0.55}{$\textsc{Diff}$}}(s,t) \right\vert ^2 ds\,dt}.
\end{equation}
The NCE can be subsequently computed as the following function of the nuclear modal fraction \cite{Karapetyan:2019ran,Karapetyan:2019fst,Karapetyan:2018oye},
\begin{equation}
\label{333}
\textsc{NCE} = - \displaystyle\iint_{\mathbb{R}^2} n_{\upsigma(s,t)} \log\left( n_{\upsigma(s,t)}\right)d s\,dt,
\end{equation}
where the CE has nat (the logarithmic natural unit of information entropy\footnote{1 nat is equivalent to the information underlying any random event to occur with a probability that equals $1/e$.}) units.

\begin{figure}[H]
 \centering
  \includegraphics[width=2.9in]{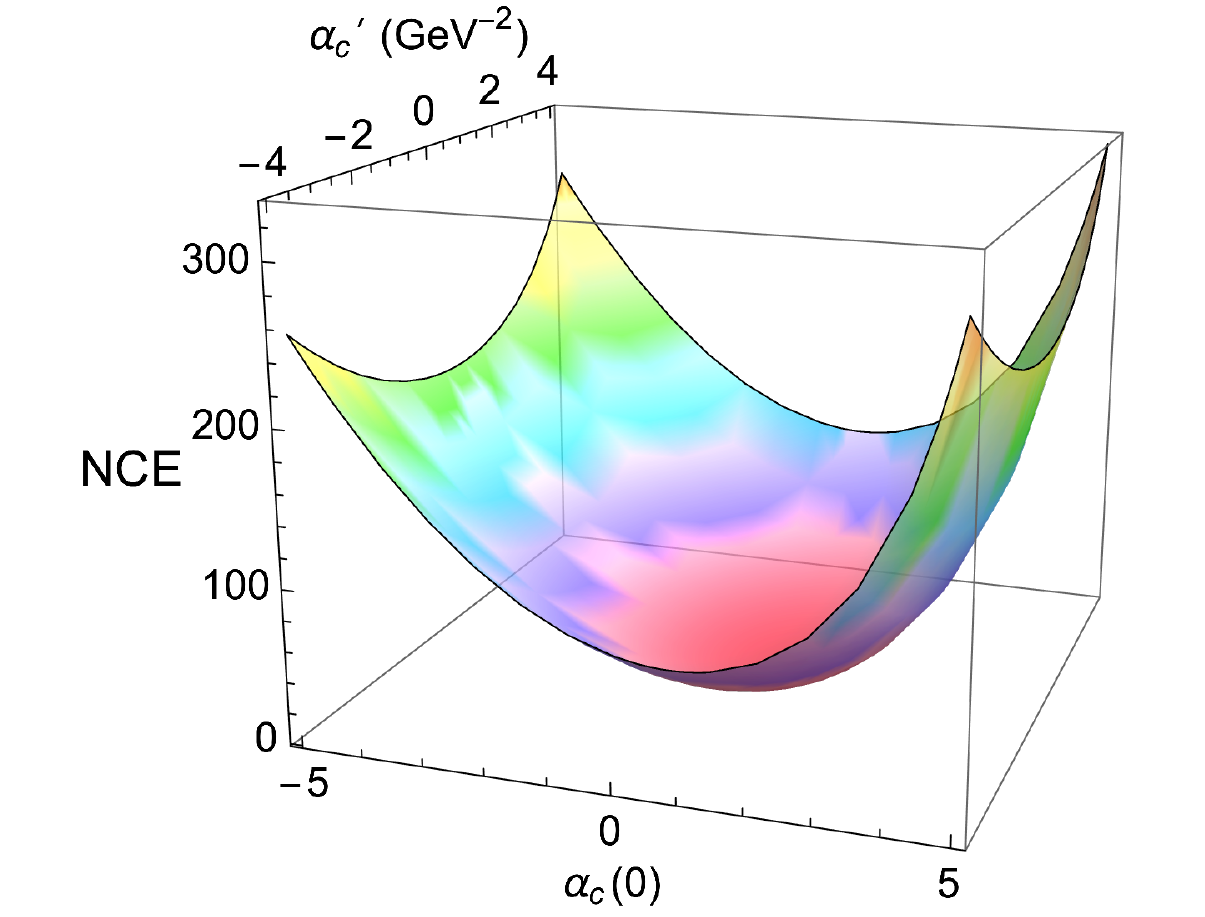}
  \caption{NCE as a function of $\upalpha_c(0)$ and $\upalpha^\prime_c$. There is a global minimum NCE$_{\scalebox{0.55}{$\textsc{min}$}}$($\upalpha^{{\scalebox{0.55}{$\textsc{min}$}}}_c(0),\upalpha^{\prime{{\scalebox{0.55}{$\textsc{min}$}}}}_c$) = 5.4482 nat, for $\upalpha^{{\scalebox{0.55}{$\textsc{min}$}}}_c(0)=1.0889$ and $\upalpha^{\prime{{\scalebox{0.55}{$\textsc{min}$}}}}_c=0.3921$ GeV${}^{-2}$, matching TOTEM data at the LHC with accuracy of 0.26\% for $\upalpha^{{\scalebox{0.55}{$\textsc{min}$}}}_c(0)$ and 0.74\% for $\upalpha^{\prime{{\scalebox{0.55}{$\textsc{min}$}}}}_c$.}
  \label{fff1}
\end{figure}

\begin{figure}[!htb]
 \centering
  \includegraphics[width=2.9in]{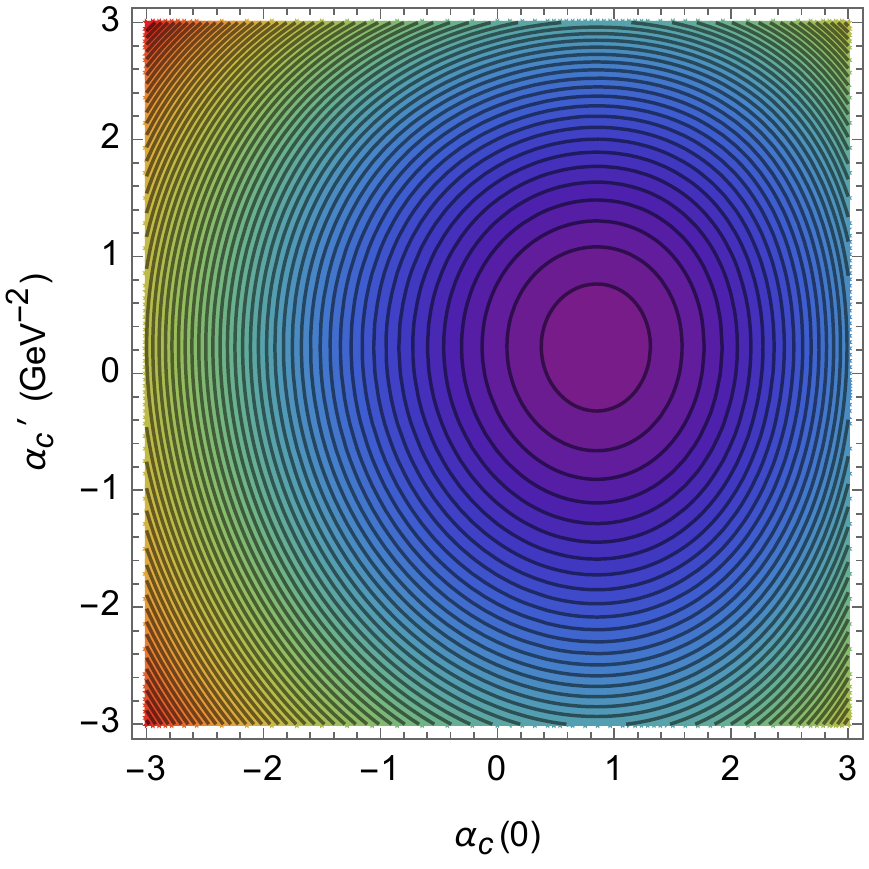}
  \caption{Contour plot of the NCE as a function of the free parameters $\upalpha_c(0)$ and $\upalpha^\prime_c$. The center of the purple ellipsis-like curve regards the values corresponding to the global minimum NCE$_{\scalebox{0.55}{$\textsc{min}$}}$($\upalpha^{{\scalebox{0.55}{$\textsc{min}$}}}_c(0),\upalpha^{\prime{{\scalebox{0.55}{$\textsc{min}$}}}}_c$) = 5.4482 nat, for $\upalpha^{{\scalebox{0.55}{$\textsc{min}$}}}_c(0)=1.0889$ and $\upalpha^{\prime{{\scalebox{0.55}{$\textsc{min}$}}}}_c=0.3921$ GeV${}^{-2}$.}
  \label{fff2}
\end{figure}

In the numerical calculations that follow, involving the NCE, Eqs. (\ref{modall}, \ref{333}) are employed, together with the expression for the
cross sections of the high energy proton-proton scattering in
a holographic QCD model given by Eq. (\ref{cs}). One must substitute Eq. (\ref{pgf}), which displays the proton gravitational form factor, into the differential cross-section (\ref{cs}). Also, let one remembers that $\uplambda = 9.95\pm0.12$ GeV${}^{-1}$, encoding the coupling between the tensors $h_{\mu\nu}$ and the QCD stress-energy-momentum tensor ${{\scalebox{0.95}{$\textsc{T}$}}}_{\mu \nu}$, has been already obtained by TOTEM data, and $\upchi^2 =1.317$ has been derived in the soft-wall model in Ref. \cite{Xie:2019soz}, fitting the data for the differential cross-section (\ref{cs}) and the total cross-section, in the range $0.546 \;{\rm TeV} \lesssim \sqrt{s}\lesssim 13$ TeV at TOTEM.
The NCE is plot in Fig. \ref{fff1}.
\noindent The global minimum
\beq
\text{NCE$_{{\scalebox{0.55}{$\textsc{min}$}}}$($\upalpha^{{\scalebox{0.55}{$\textsc{min}$}}}_c(0),\upalpha^{\prime{{\scalebox{0.55}{$\textsc{min}$}}}}_c$) = 5.4482 nat},
\eeq
at
\beq
 \text{$\upalpha^{{\scalebox{0.55}{$\textsc{min}$}}}_c(0)=1.0889$,\qquad\qquad $\upalpha^{\prime{{\scalebox{0.55}{$\textsc{min}$}}}}_c=0.3921$ GeV${}^{-2}$,}\eeq
matches TOTEM data at the LHC with accuracy of 0.26\% for $\upalpha^{{\scalebox{0.55}{$\textsc{min}$}}}_c(0)$ and 0.74\% for $\upalpha^{\prime{{\scalebox{0.55}{$\textsc{min}$}}}}_c$. The global minimum was derived through the routine {\tt NMaximize} in the {\tt Mathematica} package, taking into account Eq. (\ref{333}). The range of the parameter space $(\upalpha_c(0), \upalpha^{\prime}_c)$ here analyzed suffices to cover all the experimental possibilities. Consonant with it, Ref. \cite{Xie:2019soz} also fitted the data for the differential cross-section (\ref{cs}) and the total cross-section, in the range $0.546 \;{\rm TeV} \lesssim \sqrt{s}\lesssim 13$ TeV at TOTEM, within the AdS/QCD soft-wall. The fitting results in Ref. \cite{Xie:2019soz} yielded $\upalpha^{{\scalebox{0.55}{$\textsc{min}$}}}_c(0) = 1.086\pm0.002$ and $\upalpha^{\prime{{\scalebox{0.55}{$\textsc{min}$}}}}_c=0.395\pm0.002\,{\rm GeV}{}^{-2}$. Our results computed with in the context of the NCE corroborate with these ones, obtained in another context. 
In addition, the nuclear system here studied has higher configurational stability at the global minimum, being this mode more prevalent and dominant also from the experimental point of view. In fact, the global minimum NCE$_{\scalebox{0.55}{$\textsc{min}$}}$($\upalpha^{{\scalebox{0.55}{$\textsc{min}$}}}_c(0),\upalpha^{\prime{{\scalebox{0.55}{$\textsc{min}$}}}}_c$) = 5.4482 nat, at the point $\upalpha^{{\scalebox{0.55}{$\textsc{min}$}}}_c(0)=1.0889$ and $\upalpha^{\prime{{\scalebox{0.55}{$\textsc{min}$}}}}_c=0.3921$ GeV${}^{-2}$ corroborates the TOTEM experimental data with good accuracy.
Numerical analysis reveal that beyond the range in the plot in Fig. \ref{fff1} and \ref{fff2}, other minima of the CE lack, for all values of $\upalpha_c(0)$ and $\upalpha^\prime_c$.

The contour plot in Fig. \ref{fff2} exhibits the NCE as a function of parameters $\upalpha_c(0)$ and $\upalpha^\prime_c$, but now in the parameter space $(\upalpha_c(0), \upalpha^{\prime}_c)$.
The parameter space $(\upalpha_c(0), \upalpha^{\prime}_c)$ undergoes a
 splitting into nuclear isentropic subsectors, sorted out by gaps that are inhomogeneous, however with $\Delta [{\rm NCE}(\upalpha_c(0),\upalpha^\prime_c)] = 0.1$, in average. The nuclear configurational isentropic subsectors correspond to ellipsis-like black curves, comprising boundaries of each isentropic subsector. One can realize that the higher the major axis of the ellipsis-like curve, the lower the gap $\Delta [{\rm NCE}(\upalpha_c(0),\upalpha^\prime_c)]$ between two subsequent isentropic subdomains is.
 The outer [inner] subsectors, with hotter [colder] colors, regard higher [lower] values of the NCE. The purple central subsector encircles the global minimum NCE$_{{\scalebox{0.55}{$\textsc{min}$}}}$($\upalpha^{{\scalebox{0.55}{$\textsc{min}$}}}_c(0),\upalpha^{\prime{{\scalebox{0.55}{$\textsc{min}$}}}}_c$). 
 The further one goes away from the neighbourhood of the NCE global minimum,  the more the configurational instability increases. Table \ref{tab10} displays the values of the NCE at each extremum of the range in the parameter space $(\upalpha_c(0), \upalpha^{\prime}_c)$ here analyzed. It illustrates the huge difference between the NCE at the extrema of the parameter space range and the NCE$_{\scalebox{0.55}{$\textsc{min}$}}$($\upalpha^{{\scalebox{0.55}{$\textsc{min}$}}}_c(0),\upalpha^{\prime{{\scalebox{0.55}{$\textsc{min}$}}}}_c$) = 5.4482 nat at the global minimum. 
\begin{table}[H]
\vspace{1 mm}
\centering
\begin{tabular}{||c||c|c||}
\hline\hline
\,\diagbox{$\upalpha^\prime$ (GeV${}^{-2}$)}{$\upalpha_c(0)$}
 \,&\, $-5$ \,&\, 5 \,\\
\hline\hline
\,$-4$\,&\, 256.033 nat \,&\, 298.268 nat \,\\\hline
\,$4$\,&\, 286.557 nat \,&\, 328.757 nat \,\\\hline
\hline
\end{tabular}
\caption{\small{NCE as a function of $\upalpha_c(0)$ and $\upalpha^\prime_c$, evaluated at the extrema of the parameter space range.}}\label{tab10}
\end{table}

\section{Conclusions}
\label{con}

The NCE has been here considered as an information entropy measure in the context of AdS/QCD and the Pomeron exchange in $pp$ scattering processes. 
Using the proton gravitational form factor (\ref{pgf}) to compute the differential cross-section (\ref{cs}), the two parameters arising from the spectrum of closed strings and the Pomeron trajectory parameter of Regge theory were derived, also matching the $2^{{\scalebox{0.55}{++}}}$ glueball mass value and data of the TOTEM collaboration at the LHC. These results were achieved in the Regge limit of QCD, where the cross-sections for the proton-proton scattering process were analyzed in the context of the Reggeized glueball propagator. With it in hand, the NCE was computed and discussed. The global minimum
NCE$_{{\scalebox{0.55}{$\textsc{min}$}}}$($\upalpha^{{\scalebox{0.55}{$\textsc{min}$}}}_c(0),\upalpha^{\prime{{\scalebox{0.55}{$\textsc{min}$}}}}_c$) = 5.4482 nat, at the 2-uple $\upalpha^{{\scalebox{0.55}{$\textsc{min}$}}}_c(0)=1.0889$ and $\upalpha^{\prime{{\scalebox{0.55}{$\textsc{min}$}}}}_c=0.3921$ GeV${}^{-2}$ was derived and coincides to TOTEM data with accuracy of 0.26\% for $\upalpha^{{\scalebox{0.55}{$\textsc{min}$}}}_c(0)$ and 0.74\% for $\upalpha^{\prime{{\scalebox{0.55}{$\textsc{min}$}}}}_c$.
This 2-uple, at which the global minimum of the NCE is evaluated, corresponds to 
the point in the parameter space $(\upalpha_c(0), \upalpha^{\prime}_c)$ that drives the nuclear system towards the highest configurational stability.
This global minimum of the NCE yields the Pomeron intercept $\sim 1.084$, complying to Ref. \cite{Xie:2019soz}. The results underlying the NCE moreover point to an adequate approach to interaction processes at high energies and can carry out the analysis of other scattering processes.

\subsubsection*{Acknowledgments}
GK is grateful to the Federal University of ABC and Perimeter Institute, for the hospitality.
RdR~is grateful to The S\~ao Paulo Research Foundation -- FAPESP (Grants No. 2017/18897-8, No. 2022/01734-7, and No. 2021/01089-1) and the National Council for Scientific and Technological Development -- CNPq (Grants No. 303390/2019-0, No. 406134/2018-9, and No. 402535/2021-9), for partial financial support.
Fundação de Amparo à Pesquisa do Estado de São Paulo


\begin{thebibliography}{9}
	%%\par
	
\bibitem{Shannon:1948zz}
C.~E.~Shannon, %``A mathematical theory of communication,''
Bell Syst.\ Tech.\ J.\  \textbf{27} (1948) 379.

\bibitem{Gleiser:2011di} M. Gleiser and N. Stamatopoulos, Phys.\ Lett.\ B {\bf 713} (2012) 304 [{arXiv:1111.5597 [hep-th]}].
	%%\par
			

\bibitem{Bernardini:2016hvx} A.~E.~Bernardini and R.~da Rocha,
	%``Entropic information of dynamical AdS/QCD holographic models,''
	Phys.\ Lett.\ B {\bf 762} (2016) 107
	%%doi:10.1016/j.physletb.2016.09.023
	[{arXiv:1605.00294 [hep-th]}].	

\bibitem{Bernardini:2018uuy}
	A.~E.~Bernardini and R.~da Rocha,
	%``Informational entropic Regge trajectories of meson families in AdS/QCD,''
	Phys.\ Rev.\ D {\bf 98} (2018) 126011
	% doi:10.1103/PhysRevD.98.126011
	[arXiv:1809.10055 [hep-th]].
	%\par
	%\par

	
\bibitem{Ferreira:2019inu}
	L.~F.~Ferreira and R.~da Rocha,
	%``Pion family in AdS/QCD: the next generation from configurational entropy,''
	Phys.\ Rev.\ D {\bf 99} (2019) 086001
	%doi:10.1103/PhysRevD.99.086001
	[arXiv:1902.04534 [hep-th]].

\bibitem{Karapetyan:2018oye}
G.~Karapetyan,
%``Configurational entropy and $\rho$ and $\phi$ mesons production in QCD,''
Phys. Lett. B \textbf{781} (2018) 201
%doi:10.1016/j.physletb.2018.03.086
[arXiv:1802.09105 [nucl-th]].

\bibitem{Brodsky:2014yha} S.~J.~Brodsky, G.~F.~de Teramond, H.~G.~Dosch, J.~Erlich,
	%``Light-Front Holographic QCD and Emerging Confinement,''
	Phys.\ Rept.\  {\bf 584} (2015) 1
	% doi:10.1016/j.physrep.2015.05.001
	[{arXiv:1407.8131 [hep-ph]}].
	%\par
	
		
\bibitem{EKSS2005} J.~Erlich, E.~Katz, D.~T.~Son and M.~A.~Stephanov,
	%``QCD and a holographic model of hadrons,''
	Phys.\ Rev.\ Lett.\ \textbf{95} (2005) 261602 [{arXiv:hep-ph/0501128}].

	
\bibitem{Karch:2006pv} A.~Karch, E.~Katz, D.~T.~Son and M.~A.~Stephanov,
	%``Linear confinement and AdS/QCD,''
	Phys.\ Rev.\ D {\bf 74} (2006) 015005
	% doi:10.1103/PhysRevD.74.015005
	[{hep-ph/0602229}].
	%\par

\bibitem{Colangelo:2008us}
P.~Colangelo, F.~De Fazio, F.~Giannuzzi, F.~Jugeau and S.~Nicotri,
%``Light scalar mesons in the soft-wall model of AdS/QCD,''
Phys. Rev. D \textbf{78} (2008) 055009
%doi:10.1103/PhysRevD.78.055009
[arXiv:0807.1054 [hep-ph]].

\bibitem{Branz:2010ub}
T.~Branz, T.~Gutsche, V.~Lyubovitskij, I.~Schmidt, A.~Vega,
%``Light and heavy mesons in a soft-wall holographic approach,''
Phys. Rev. D \textbf{82} (2010) 074022
[arXiv:1008.0268 [hep-ph]].

\bibitem{daRocha:2021ntm}
R.~da Rocha,
%``Information entropy in AdS/QCD: Mass spectroscopy of isovector mesons,''
Phys. Rev. D \textbf{103} (2021) 106027
%doi:10.1103/PhysRevD.103.106027
[arXiv:2103.03924 [hep-ph]].
	
	
\bibitem{daRocha:2021imz}
R.~da Rocha,
%``Deploying heavier $\upeta$ meson states: Configurational entropy hybridizing AdS/QCD,''
Phys. Lett. B \textbf{814} (2021) 136112
%doi:10.1016/j.physletb.2021.136112
[arXiv:2101.03602 [hep-th]].
	
\bibitem{Ferreira:2019nkz}
	L.~F.~Ferreira and R.~da Rocha,
	%\emph{Tensor mesons, AdS/QCD and information,}
	Phys. Rev. D \textbf{101} (2020) 106002
	[arXiv:1907.11809 [hep-th]].

	
\bibitem{Barbosa-Cendejas:2018mng} N.~Barbosa-Cendejas, R.~Cartas-Fuentevilla, A.~Herrera-Aguilar, R.~R.~Mora-Luna and R.~da Rocha,
%	``Dynamical tachyonic AdS/QCD and information entropy,''
	Phys.\ Lett.\ B {\bf 782} (2018) 607
	% doi:10.1016/j.physletb.2018.06.007
	[{arXiv:1805.04485 [hep-th]}].

	
\bibitem{Braga:2018fyc} N.~R.~F.~Braga, L.~F.~Ferreira and R.~da Rocha,
	%\emph{Thermal dissociation of heavy mesons and configurational entropy,}
	Phys.\ Lett.\ B {\bf 787} (2018) 16
	[{arXiv:1808.10499 [hep-ph]}].

\bibitem{Karapetyan:2018yhm}
	G.~Karapetyan,
	%``The nuclear configurational entropy approach to dynamical QCD effects,''
	Phys.\ Lett.\ B {\bf 786} (2018) 418 [arXiv:1807.04540 [nucl-th]].
	%\par

	
\bibitem{Karapetyan:2019ran}
G.~Karapetyan,
%``Hadron multiplicity calculation: a configurational entropy approach to the saturation scale in QCD,''
EPL \textbf{129} (2020)  18002
%doi:10.1209/0295-5075/129/18002
[arXiv:1912.10071 [hep-ph]].
	
\bibitem{Ma:2018wtw}
	C.~W.~Ma and Y.~G.~Ma,
	%``Shannon Information Entropy in Heavy-ion Collisions,''
	Prog.\ Part.\ Nucl.\ Phys.\  {\bf 99} (2018) 120
	%doi:10.1016/j.ppnp.2018.01.002
	[arXiv:1801.02192 [nucl-th]].

\bibitem{Bernardini:2016qit} A.~E.~Bernardini, N.~R.~F.~Braga and R.~da Rocha,
	%``Configurational entropy of glueball states,''
	Phys.\ Lett.\ B {\bf 765}  (2017) 81  [{arXiv:1609.01258 [hep-th]}].
	%\par
	
	
\bibitem{Braga:2017fsb} N.~R.~F.~Braga and R.~da Rocha,
%``AdS/QCD duality and the quarkonia holographic information entropy,''
	Phys.\ Lett.\ B {\bf 776} (2018) 78 [{arXiv:1710.07383 [hep-th]}].
	%\par
	
\bibitem{Colangelo:2018mrt}
	P.~Colangelo and F.~Loparco,
	%``Configurational Entropy can disentangle conventional hadrons from exotica,''
	Phys.\ Lett.\ B {\bf 788} (2019) 500
	%doi:10.1016/j.physletb.2018.11.053
	[arXiv:1811.05272 [hep-ph]].
	
\bibitem{Ferreira:2020iry}
L.~F.~Ferreira and R.~da Rocha,
%``Nucleons and higher spin baryon resonances: An AdS/QCD configurational entropic incursion,''
Phys. Rev. D \textbf{101} (2020) 106002
%doi:10.1103/PhysRevD.101.106002
[arXiv:2004.04551 [hep-th]].
	

\bibitem{MarinhoRodrigues:2020yzh}
D.~Marinho Rodrigues and R.~da Rocha,
%``Odd-spin glueballs, AdS/QCD and information entropy,''
Phys. Lett. B \textbf{811} (2020) 135943
%doi:10.1016/j.physletb.2020.135943
[arXiv:2009.01890 [hep-th]].
	
\bibitem{pdg} P. A. Zyla et al. (Particle Data Group), Prog. Theor. Exp. Phys. {\bf 2020} (2020) 083C01.

\bibitem{FolcoCapossoli:2016uns}
E.~Folco Capossoli, D.~Li,  H.~Boschi-Filho,
%``Dynamical corrections to the anomalous holographic soft-wall model: the pomeron and the odderon,''
Eur. Phys. J. C \textbf{76} (2016) 320
%doi:10.1140/epjc/s10052-016-4171-0
[arXiv:1604.01647 [hep-ph]].

\bibitem{Karapetyan:2021vyh}
G.~Karapetyan,
%``Total hadronic and photonic cross sections and the nuclear configurational entropy concept,''
Eur. Phys. J. Plus \textbf{136} (2021) 1012
%doi:10.1140/epjp/s13360-021-02017-3
[arXiv:2105.07546 [hep-ph]].
	
\bibitem{Karapetyan:2021ufz}
G.~Karapetyan and R.~da Rocha,
%``Configurational entropy of heavy-quark QCD exotica,''
Eur. Phys. J. Plus \textbf{136} (2021) 993
%doi:10.1140/epjp/s13360-021-01942-7
[arXiv:2103.10863 [hep-ph]].
	
\bibitem{Karapetyan:2020yhs}
G.~Karapetyan,
%``Baryon production probability via the nuclear configurational entropy,''
Eur. Phys. J. Plus \textbf{136} (2021) 122
%doi:10.1140/epjp/s13360-021-01076-w
[arXiv:2003.08994 [hep-ph]].
	
\bibitem{Gutsche:2016wix}
T.~Gutsche, S.~Kuleshov, V.~E.~Lyubovitskij and I.~T.~Obukhovsky,
%``Search for the glueball content of hadrons in $\gamma p$ interactions at GlueX,''
Phys. Rev. D \textbf{94} (2016) 034010
%doi:10.1103/PhysRevD.94.034010
[arXiv:1605.01035 [hep-ph]].

\bibitem{Sergeenko:2011kf}
	M.~Sergeenko,
	%``Glueballs and the Pomeron,''
	EPL \textbf{89} (2010) 11001
	%doi:10.1209/0295-5075/89/11001
	[arXiv:1107.1671 [hep-ph]].


\bibitem{Xie:2019soz}
W.~Xie, A.~Watanabe and M.~Huang,
%``Elastic proton-proton scattering at LHC energies in holographic QCD,''
JHEP \textbf{10} (2019) 053
[arXiv:1901.09564 [hep-ph]].

\bibitem{TOTEM:2017asr}
G.~Antchev \textit{et al.} [TOTEM],
%``First measurement of elastic, inelastic and total cross-section at $\sqrt{s}=13$ TeV by TOTEM and overview of cross-section data at LHC energies,''
Eur. Phys. J. C \textbf{79} (2019) 103
%doi:10.1140/epjc/s10052-019-6567-0
[arXiv:1712.06153 [hep-ex]].

\bibitem{TOTEM:2012oyl}
G.~Antchev \textit{et al.} [TOTEM],
%``Luminosity-Independent Measurement of the Proton-Proton Total Cross Section at $\sqrt{s}=8$  TeV,''
Phys. Rev. Lett. \textbf{111} (2013) 012001.


\bibitem{FolcoCapossoli:2019imm}
E.~Folco Capossoli, M.~A.~Mart\'\i{}n Contreras, D.~Li, A.~Vega and H.~Boschi-Filho,
%``Hadronic spectra from deformed AdS backgrounds,''
Chin. Phys. C \textbf{44} (2020) 064104
%doi:10.1088/1674-1137/44/6/064104
[arXiv:1903.06269 [hep-ph]].

\bibitem{Rodrigues:2016cdb}
D.~M.~Rodrigues, E.~F. Capossoli, H.~Boschi-Filho,
%``Twist Two Operator Approach for Even Spin Glueball Masses and Pomeron Regge Trajectory from the Hardwall Model,''
Phys. Rev. D \textbf{95} (2017) 076011
%doi:10.1103/PhysRevD.95.076011
[arXiv:1611.03820 [hep-th]].

\bibitem{FolcoCapossoli:2015jnm}
E.~Folco Capossoli and H.~Boschi-Filho,
%``Glueball spectra and Regge trajectories from a modified holographic softwall model,''
Phys. Lett. B \textbf{753} (2016) 419
%doi:10.1016/j.physletb.2015.12.034
[arXiv:1510.03372 [hep-ph]].

\bibitem{Boschi-Filho:2005xct}
H.~Boschi-Filho, N.~R.~F.~Braga, H.~L.~Carrion,
%``Glueball Regge trajectories from gauge/string duality and the Pomeron,''
Phys. Rev. D \textbf{73} (2006) 047901
%doi:10.1103/PhysRevD.73.047901
[arXiv:hep-th/0507063].

\bibitem{Ballon-Bayona:2017vlm}
A.~Ballon-Bayona, R.~C.  Quevedo and M.~S.~Costa,
%``Unity of pomerons from gauge/string duality,''
JHEP \textbf{08} (2017) 085
%doi:10.1007/JHEP08(2017)085
[arXiv:1704.08280 [hep-ph]].

\bibitem{Hu:2017iix}
Z.~Hu, B.~Maddock and N.~Mann,
%``A Second Look at String-Inspired Models for Proton-Proton Scattering via Pomeron Exchange,''
JHEP \textbf{08} (2018) 093
%doi:10.1007/JHEP08(2018)093
[arXiv:1710.02463 [hep-ph]].

\bibitem{Domokos:2009hm}
S.~K.~Domokos, J.~A.~Harvey and N.~Mann,
%``The Pomeron contribution to p p and p anti-p scattering in AdS/QCD,''
Phys. Rev. D \textbf{80} (2009) 126015
%doi:10.1103/PhysRevD.80.126015
[arXiv:0907.1084 [hep-ph]].

\bibitem{Karapetyan:2019fst}
G.~Karapetyan,
%``Nuclear configurational entropy of the energy-energy correlation in $e^+e^-$ annihilation processes,''
EPL \textbf{125} (2019) 58001
%doi:10.1209/0295-5075/125/58001
[arXiv:1901.05349 [hep-ph]].

\end{thebibliography}
\end{document}